\documentclass[apjl,twocolumn]{emulateapj_mod}

\def\gtsima{$\; \buildrel > \over \sim \;$}
\def\ltsima{$\; \buildrel < \over \sim \;$}
\def\prosima{$\; \buildrel \propto \over \sim \;$}
\def\gsim{\lower.5ex\hbox{\gtsima}}
\def\lsim{\lower.5ex\hbox{\ltsima}}
\def\simgt{\lower.5ex\hbox{\gtsima}}
\def\simlt{\lower.5ex\hbox{\ltsima}}
\def\simpr{\lower.5ex\hbox{\prosima}}

\def\h1{$h^{-1}$}
\def\eeq{\end{equation}}
\def\beq{\begin{equation}}
\def\24mu{24\,$\mu{\rm m}$}
\def\70mu{70\,$\mu{\rm m}$}
\def\8mu{8\,$\mu{\rm m}$}

\newcommand{\oiii}{[O\,{\sc iii}$]\lambda$500.7}

\newcommand{\rga}{{\rm RG\,J163655}}
\newcommand{\rgb}{{\rm RG\,J131236}}
\newcommand{\rgc}{{\rm RG\,J123711}}
\newcommand{\pdbishort}{{\rm PdBI }}
\newcommand{\pdbilong}{{\rm IRAM -- Plateau de Bure Interferometer}}

\shorttitle{CO in $z\sim2$ SFRGs}
\shortauthors{S.\ C.\ Chapman et al.}

\journalinfo{Astrophysical Journal in press}
\begin{document}

\title{Interferometric CO Observations of submillimeter-faint, radio-selected starburst galaxies at  $\lowercase{z}\sim2$}

 \author{S.\ C.\ Chapman,\altaffilmark{1,2}
R.\ Neri,\altaffilmark{3}
F.\ Bertoldi,\altaffilmark{4}
Ian Smail,\altaffilmark{5}
T.\ R.\ Greve,\altaffilmark{6}
D.\ Trethewey,\altaffilmark{1}  
A.\ W.\ Blain,\altaffilmark{7}\\
P.\ Cox,\altaffilmark{3}
R.\ Genzel,\altaffilmark{8}
R.\ J.\ Ivison,\altaffilmark{9,10}
A.\ Kovacs,\altaffilmark{4}
A.\ Omont\altaffilmark{11} and
A.\ M.\ Swinbank\altaffilmark{5}
}

\altaffiltext{1}{Institute of Astronomy, Madingley Road, Cambridge, CB3\,0HA, UK.}
\altaffiltext{2}{University of Victoria, Victoria, BC, V8W\,3P6, Canada.}
\altaffiltext{3}{Institut de Radio Astronomie Millim\'etrique (IRAM),
                 St Martin d'H\`{e}res, France.}
\altaffiltext{4}{Argelander Institut fur Astronomie,
Universitat Bonn, Auf dem Hugel 71, 53121 Bonn}
\altaffiltext{5}{Institute for Computational Cosmology, Durham University,
                 South Road, Durham DH1\,3LE, UK.}
\altaffiltext{6}{Astronomy Department, Max- Planck Institut f\"{u}r Astronomie, Koningstuhl-17, D-69117, Heidelberg, Germany}
\altaffiltext{7}{California Institute of Technology, Pasadena, CA\,91125.}
\altaffiltext{8}{Max-Planck Institut f\"{u}r extraterrestrische Physik (MPE),
                 Garching, Germany.}
\altaffiltext{9}{UK Astronomy Technology Centre, Royal Observatory,
                 Blackford Hill, Edinburgh EH9\,3HJ, UK.}
\altaffiltext{10}{Institute for Astronomy, University of Edinburgh, Royal Observatory,
                 Blackford Hill, Edinburgh EH9\,3HJ, UK.}
\altaffiltext{11}{Institut d'Astrophysique de Paris, CNRS, Universit\'{e} de Paris,
                  Paris, France.}

\begin{abstract}
High-redshift, dust-obscured galaxies -- selected to be luminous in
the radio but relatively faint at 850\,$\mu$m -- appear to represent a
different population from the ultra-luminous
submillimeter- (submm-) bright population. They may be star-forming
galaxies with hotter dust temperatures or they may have lower
far-infrared luminosities and larger contributions from obscured
active galactic nuclei (AGN).  Here we present observations of three 
$z\sim2$ examples of this population, which we term {\it submm-faint radio galaxies -- SFRGs}
in CO(3--2) using the IRAM Plateau de Bure
Interferometer to study their gas and dynamical properties. 
We estimate the molecular gas mass in each of the
three SFRGs ($8.3\times10^{9}$M$_\odot$, $<5.6\times10^{9}$M$_\odot$
and $15.4\times10^{9}$M$_\odot$, respectively) and, in the case of
\rga, a dynamical mass by measurement of the width of the CO(3--2)
line ($8\times10^{10} \csc^2i$\,M$_\odot$).  While these gas masses
are substantial, on average they are 4$\times$ lower than
submm-selected galaxies (SMGs). Radio-inferred star formation rates ($<{\rm SFR_{\rm
radio}}>=970$\,M$_{\odot}\,$yr$^{-1}$) suggest much higher
star-formation efficiencies than are found for SMGs, and shorter gas
depletion time scales ($\sim$11\,Myr), much shorter than the time
required to form their current stellar masses ($\sim$160\,Myr;
$\sim$10$^{11}$\,M$_\odot$). By contrast, SFRs may be overestimated by
factors of a few, bringing the efficiencies in line with those
typically measured for other ultraluminous star-forming galaxies and
suggesting SFRGs are more like ultraviolet- (UV-)selected star-forming
galaxies with enhanced radio emission.  A tentative detection of \rga\
at 350\,$\mu$m suggests hotter dust temperatures -- and thus similar
gas-to-dust mass fractions -- as the SMGs.  
We conclude that SFRGs' radio luminosities are larger than would naturally
scale from local ULIRGs given their gas masses or gas fractions.
\end{abstract}

\section{Introduction}

Submm surveys have provided an efficient probe of star-formation
activity in ultraluminous infrared (IR) galaxies (ULIRGs,
$>$10$^{12}$\,L$_\odot$) in the distant Universe
(e.g., Smail,
Ivison \& Blain 1997; Hughes et al.\ 1998; Barger et al.\ 1998), with bright submm
emission providing unambiguous evidence of massive quantities of dust,
heated predominantly by young stars rather than AGN (e.g.,  Chapman
et al.\ 2003a; Alexander et al.\ 2005; Menendez-Delmestre et al.\ 2007; Valiante et al.\ 2007; Pope et al.\ 2008).  Before the availability of
the Atacama Large Millimeter Array (ALMA), confusion will continue to
limit the sensitivity of current submm surveys. As a result, many
ULIRGs fall below the detection limits due to variations in their
spectral energy distributions (SEDs) -- usually parameterized in terms
of dust temperature ($T_{\rm d}$) -- meaning that entire populations
of star-forming galaxies may have been missed by submm surveys. 

For a fixed far-infrared luminosity (FIR), a galaxy with a higher $T_{\rm d}$ will be weaker in the submm at 850$\mu$m than a galaxy with a lower $T_{\rm d}$. Specifically, raising $T_{\rm d}$ from the canonical
$\sim$35~K for SMGs to 45~K will result in a factor $\sim$10$\times$ drop in 850$\mu$m flux density
(Blain 1999; Chapman et al.\ 2004).
These
galaxies should, though, be accessible in the radio waveband,
regardless of their specific SEDs, since the radio correlates with the integrated FIR emission 
(Helou et al.\ 1985) with a small $\sim0.2$dex dispersion and no observable dependence on SED type.
However,  there is potential for
large AGN contaminations in the radio, as has often been the case with mid-IR
selection of $z>1$ ULIRGs (e.g., Houck et al.\ 2005; Yan et al.\ 2005,
2007; Sajina et al.\ 2007; Weedman et al.\ 2006a, 2006b; Desai et al.\
2006) and the facilities required to provide the high-resolution,
multi-frequency radio data needed to decontaminate the samples (e.g.,
Ivison et al.\ 2007a) are not yet available.

Substantial populations of apparently star-forming galaxies at
$z\sim2$ have been uncovered through deep 1.4-GHz radio continuum
observations, many of which are not detected at submm wavelengths with
the current generation of instruments (Barger et al.\ 2000; Chapman et
al.\ 2001, 2003b, 2004a -- hereafter C04).  These galaxies are
luminous in the radio and spectroscopy suggests that star formation is
powering their bolometric output (there is little or no sign of
high-ionization emission lines, characteristic of AGN in their
UV/optical spectra).  These galaxies could, in principle, span a range
in properties from deeply obscured AGN to far-IR-luminous
starbursts. In the latter case, one would expect a different SED from
a typical SMG -- a higher $T_{\rm d}$ for instance.  These {\it
submm-faint radio sources} have a large volume density at $z\sim2$,
even larger than the SMGs (Haarsma et al.\ 1998; Richards et al.\
1999; Chapman et al.\ 2003a, C04; Barger et al.\ 2007). There are
$\rho = 2 \times 10^{-5}$\,Mpc$^{-3}$ radio sources with L$_{\rm 1.4
GHz}>10^{31}$ ergs\,s$^{-1}$\,Hz$^{-1}$ at $z\sim2$ compared with
$\rho = (6.2\pm2.3) \times 10^{-6}$~Mpc$^{-3}$ for SMGs brighter than
5\,mJy at 850\,$\mu$m at the same epoch (Chapman et al.\ 2003b, 2005).
As essentially all of these SMGs form a subset of these radio sources
(Chapman et al.\ 2005; Pope et al.\ 2006; cf.\ Ivison et al.\ 2002)
this implies $\sim14\times10^{-6}$~Mpc$^{-3}$ luminous radio sources
remain undetected at submm wavelengths. Understanding the exact
properties of these galaxies is therefore of great importance. If they
are all forming stars at the rates implied by their radio
luminosities, they would triple the observed SFR density (SFRD) at
$z\sim2$. By contrast, if their radio luminosity comes from a mix of
star formation and AGN, they have less impact on the global SFRD but
they increase the highly obscured AGN fraction at these epochs (e.g.,
Daddi et al.\ 2007a; Casey et al.\ 2008) and contribute substantially
to black hole growth.

Together with other observations, the redshifted cooling emission
lines of CO allow us to assess and compare the energy source of SFRGs
with that of SMGs and other distant star-forming galaxies via
measurements of their gas and dynamical masses. In this paper, we
present the results of a pilot study with the \pdbilong\ to detect
molecular gas in SFRGs through the rotational CO(3--2) line emission.
In \S~2 we describe the sample properties and observations both with \pdbishort\ and other facilities.
Section \S~3 presents the CO(3--2) detections and limits obtained from the \pdbishort\ observations, \S~4.1 estimates gas properties, star-formation rates and efficiencies, and \S~4.2 compares the SFRGs to other galaxy populations. Finally \S~5 discusses the results and places them in a broader galaxy evolution context.
Throughout we assume a cosmology with $h=0.7, \Omega_\Lambda = 0.72, \Omega_M = 0.28$ (e.g., Hinshaw et al.\ 2008).

\begin{figure}
\centering
\includegraphics[width=5.1cm,angle=0]{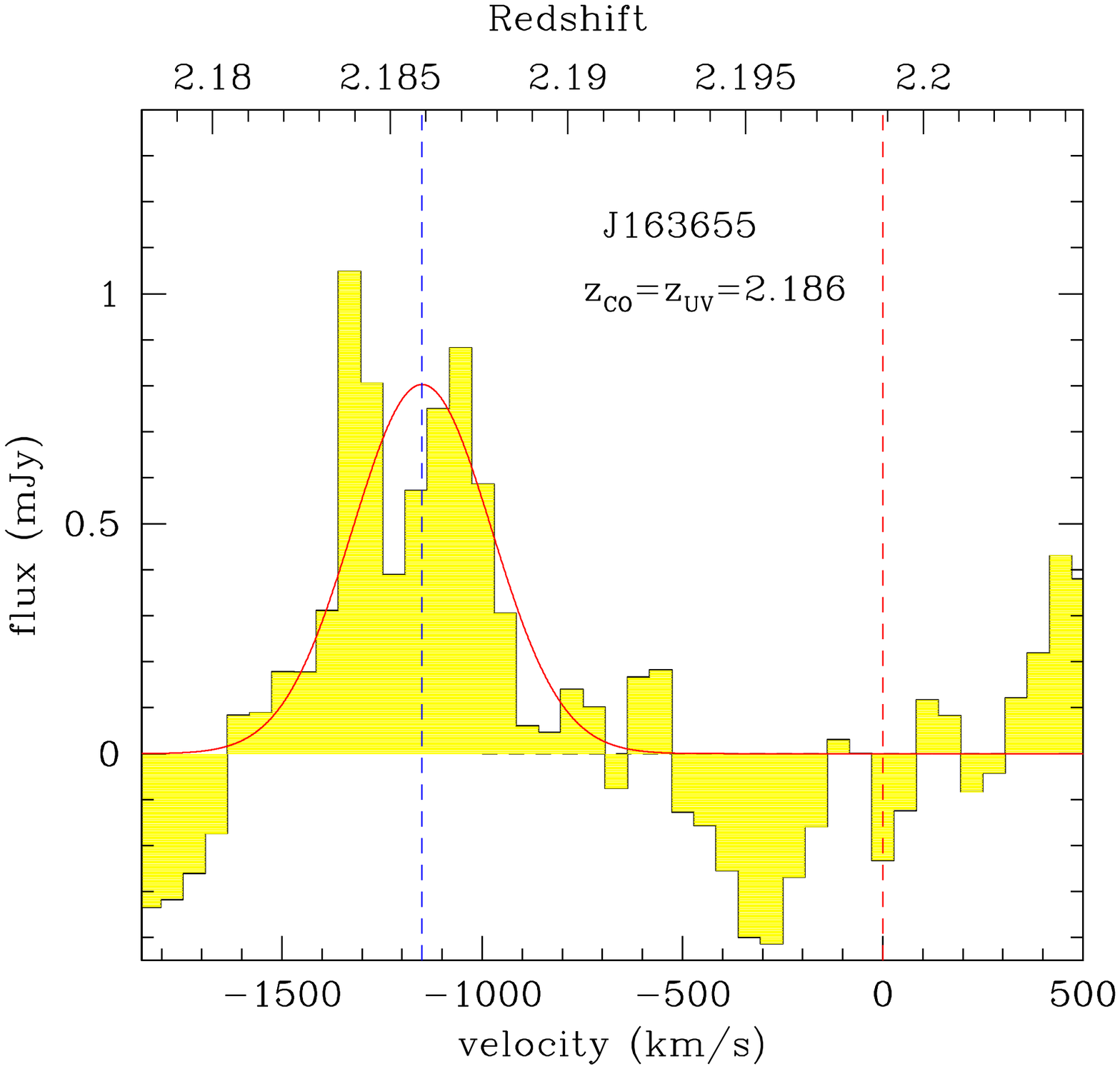} 
\includegraphics[width=4.7cm,angle=-90]{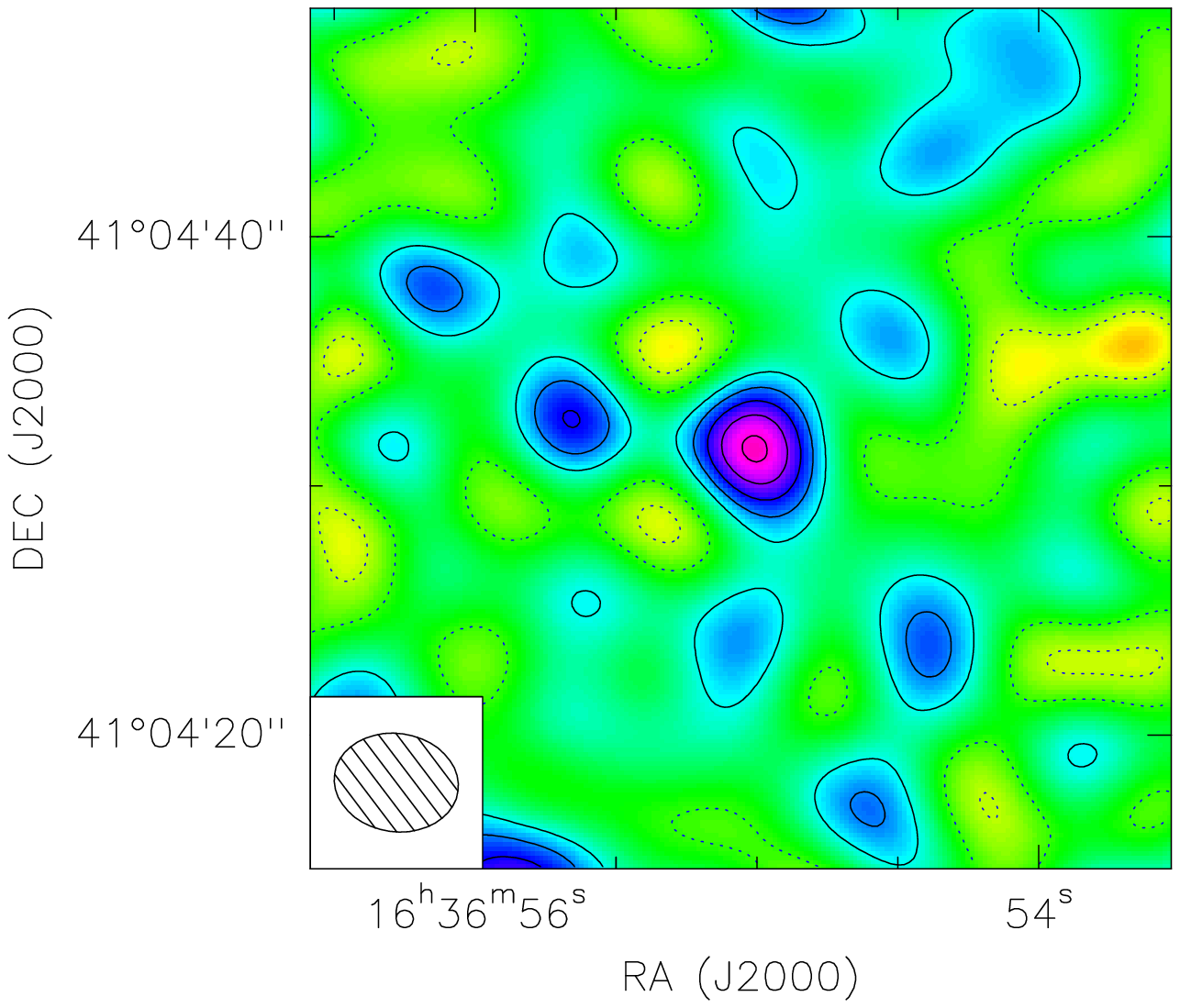} 
\includegraphics[width=5.1cm,angle=0]{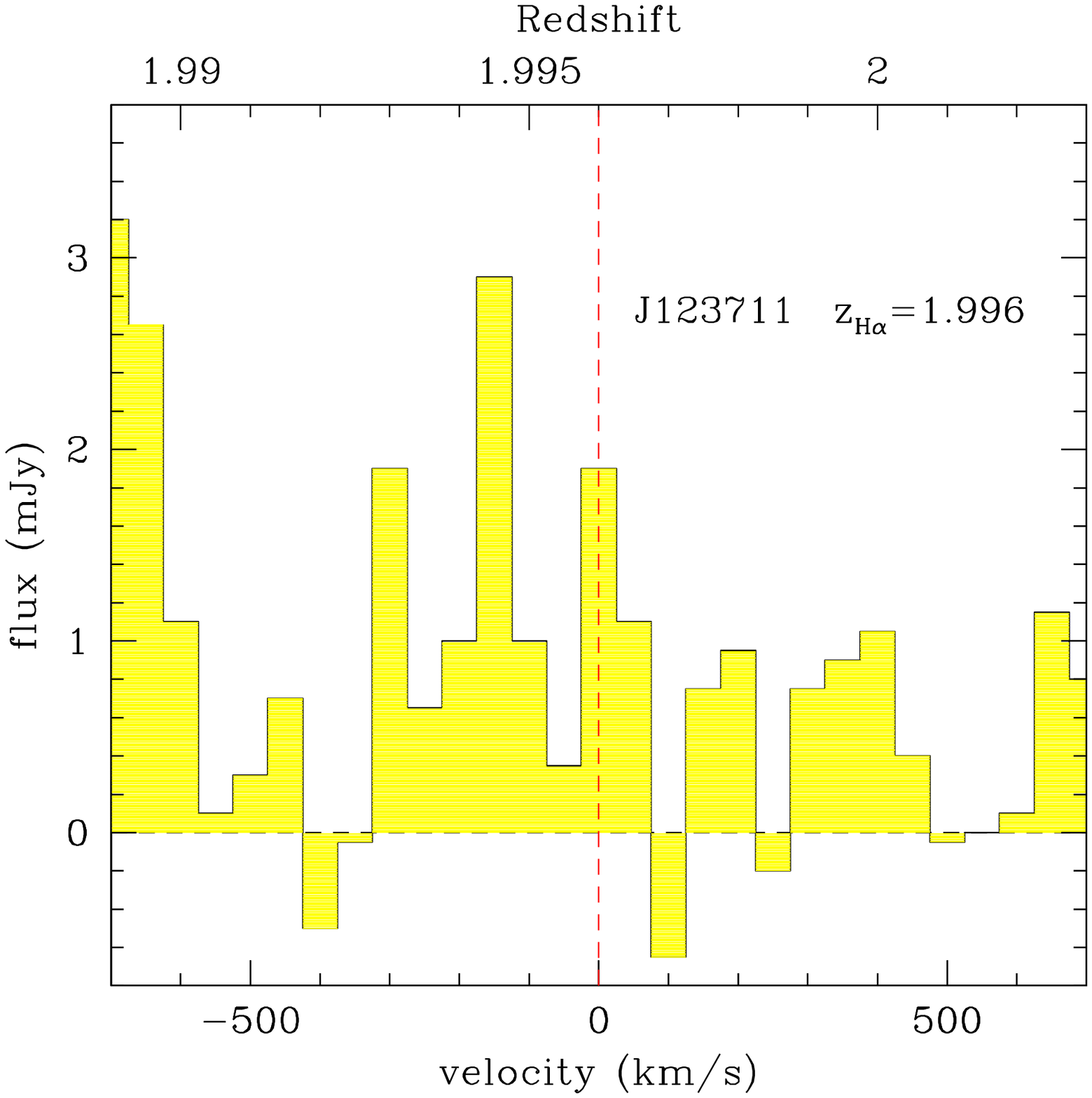} 
\includegraphics[width=4.7cm,angle=-90]{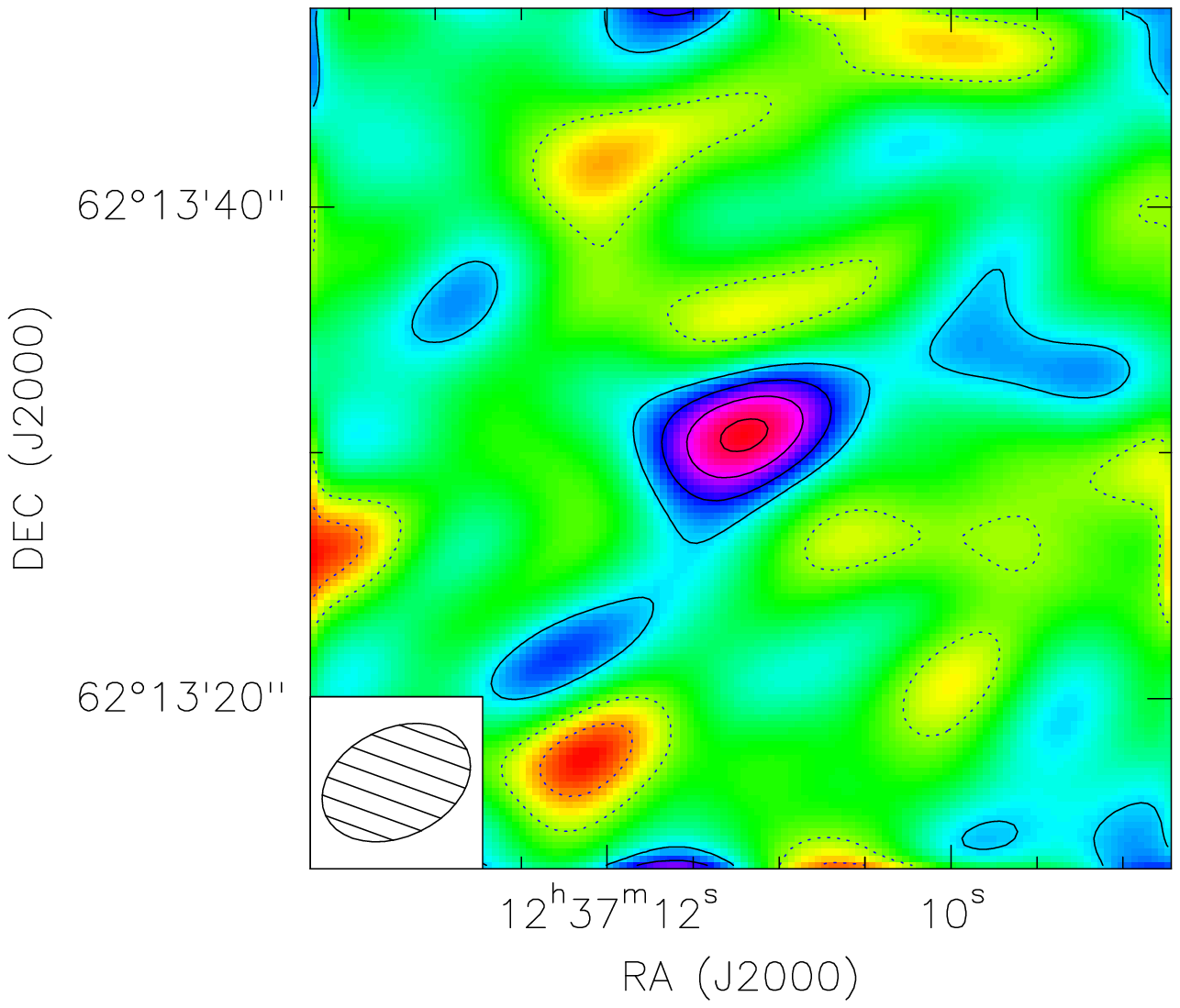} 
\vskip0.4cm
\caption{{\bf top panels:} CO(3--2) spectra for the two candidate detections.
 The spectra are shown smoothed with a 50~km/s boxcar filter, and with respect to the zero velocity offsets defined from the H$\alpha$ emission line redshift
(red dashed line). The best-fit Gaussian profile is
shown for the emission line in \rga, along with the Uv-Inferred redshift from inter-stellar absorption
lines (blue dashed line). 
{\bf bottom panels:} velocity-averaged
spatial maps of CO emission, from $-$1500 to $-$800\,km\,s$^{-1}$ (\rga) where contours are from $-$1 to 5$\sigma$ in steps of $\sigma$
(0.05\,mJy\,beam$^{-1}$). \rgc\ does not represent a formal CO detection. We measure a significance of 3.2$\sigma$ integrating over the full band).
The field of view, on a side, is 35\arcsec, with the size of the beam shown to the
lower left.  Both CO emitters lie exactly at the radio source position to within 1\arcsec.}
\label{Pic}
\end{figure}

\section{Sample Properties and Observations}

Our sample is drawn from an expansion of the C04 submm-faint,
radio-selected galaxy (SFRG) program, with galaxies drawn from several
deep radio survey fields with typical sensitivity limits of
$\sigma=4-8$\,$\mu$Jy (e.g., Biggs \& Ivison 2006). In the submm, the
survey fields are imaged to a typical depth of $\sigma_{\rm 850 \mu
m}\sim1-2$\,mJy (e.g.,  Scott et al.\ 2002, Borys et al.\ 2003).  We selected sources with
redshifts, radio luminosities, and submm limits typical of the population ($<z>=2.1$,
$<L_{1.4 GHz}>=2\times10^{31}$~ergs\,s$^{-1}$\,Hz$^{-1}$, $L_{850 \mu m}<1\times10^{31}$~ergs\,s$^{-1}$\,Hz$^{-1}$ ($<2$~mJy for $z\sim2$) at the $\sim2\sigma$ level, lying within
$\pm1\sigma$ of the median SFRG in (C04), and observing \rga\ and \rgb\
based on suitability of RA and confidence in the optical spectroscopic
redshifts.  A third source from our sample, \rgc, was observed
previously with \pdbishort in the same field as an SMG in the program described by Neri et al.\
(2003) and Greve et al.\ (2005), and we include
this object here.  We note that while the neighboring SMG 
(SMM\,J123712.0+621326) lies only 8\arcsec\ to the south-east, we are confident
that \rgc\ is not a luminous submm emitter.  Firstly, SMM\,J123712,
has a strong CO line detection (Smail et al., in preparation) with a
$\sim5\sigma$ detection $S_{\rm CO(3-2)} = 1.2$\,Jy\,km\,s$^{-1}$,
comparable to typical SMGs from Greve et al.\ (2005).  Secondly, the
850-$\mu$m emission peak (with a R=7$''$ beam) in our SCUBA map is
centered on the radio source position, whereas no significant peak is
observed at the position of \rgc. Removing a 850-$\mu$m point source
from the position of SMM\,J123712 reveals an even lower 850-$\mu$m
flux density ($0.1\pm1.2$\,mJy) at the position of \rgc\ than the
$2.2\pm1.2$\,mJy conservatively adopted for our calculations
(Table~2). Importantly, the small 850-$\mu$m/1.4-GHz flux ratio for
this source is clearly comparable to SFRGs and not to the typical
radio-detected SMGs in Chapman et al.\ (2005) or Ivison et al.\
(2007b). The properties of these SFRGs are listed in Tables~1 and 2,
and displayed in Figs~1 and 2.

\subsection{\pdbishort observations}

\rga\ and \rgb\ were observed in their redshifted CO(3--2) lines and
in the continuum at $\sim$108\,GHz using the newly refurbished PdBI
receivers for 11.5 and 4.8\,hr, respectively.  Observations were made
in D configuration on 2007 January 24, April 28, May 08 and June\ 03,
with good atmospheric phase stability (seeing, 0.7--1.4\arcsec) and
reasonable transparency (0.5\,mm of precipitable water vapor).  For
RG\,J163655, the H$\alpha$ redshift showed a considerable offset from
that inferred from UV absorption and emission lines. While this was
conceivably due to a large velocity starburst wind
($\sim$1300\,km\,s$^{-1}$), we considered the possibility that one
redshift might have calibration or resolution problems.  We therefore
observed \rga\ split over two slightly offset frequency settings to
span all measured redshifts. The overall flux scale for each observing
epoch was calibrated using a variety of sources. In each observing
epoch between three and six sources were used.  The visibilities were
resampled to a velocity resolution of 55\,km\,s$^{-1}$ (20\,MHz)
providing 1$\sigma$ line sensitivities of 1.6\,mJy\,beam$^{-1}$.  The
corresponding synthesized beam, adopting natural weighting, was similar for both sources,
5.0\arcsec\ by 4.0\arcsec\ at PA $\sim$80 degrees, east of north.
Observations of \rgc\ proceeded similarly to those described by Greve et al.\ (2005).
The \pdbishort\ data for all three SFRGs were calibrated, mapped and
analyzed using the GILDAS software package. 
The CO(3--2) spectra and images of the two candidate SFRG detections,
\rga\ and \rgc, are shown in Fig.~1.

\subsection{Other Observations}

{\em Spitzer} observations of the SFRGs in this paper were taken with
the Infrared Array Camera (IRAC) and the Multi-band Imaging Photometer
(MIPS) through various GTO and Legacy programs. 

\begin{figure}
\vskip-4cm
\centering
\includegraphics[width=9cm]{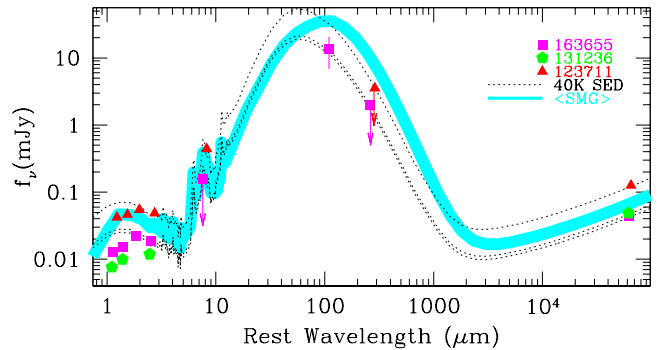} 
\caption{SEDs for \rga, \rgb\ and \rgc. A local galaxy SED with
$T_{\rm d}=40$\,K (from Dale et al.\ 2003) is shown, normalized to the
radio point in each case. An average fit to the SEDs of representative
SMGs is shown as a heavy line (Chapman et al.\ 2005; Kovacs et al.\
2006; Hainline et al.\ 2008).}
\label{Spe}%
\end{figure}

{\bf \rga:}
In addition to the SCUBA 850-$\mu$m photometry, observations 
were obtained at 350\,$\mu$m with the SHARC2 camera (Dowell et al.\
2005) on the Caltech Submillimeter Observatory as part of the imaging
campaign of Kovacs et al.\ (2006), where observational details can be
found. 
A near-IR spectrum from UKIRT/UIST covering the
H$\alpha$/[N\,{\sc ii}] region (Swinbank et al.\ 2006) finds a large
H$\alpha$/[N\,{\sc ii}] ratio suggestive of a relatively low
metallicity, [Fe/H]~$\sim$~$-$0.9, and no strong AGN component. While
the \oiii\ region was also covered with the instrument, we did not
clearly detect any emission lines, setting a limit on the \oiii\ line 
of $3.75 \times 10^{-17}$\,W\,m$^{-2}$ 
again suggesting that an AGN does not dominate the energetics of this
galaxy.  The line width of the H$\alpha$ emission (FWHM$_{\rm rest}$)
is 420$\pm$100\,km\,s$^{-1}$ and the integrated line flux of $5.0
\times 10^{-19}$\,W\,m$^{-2}$ then suggests a SFR uncorrected for dust
extinction of 150$\pm$50\,M$_\odot$\,yr$^{-1}$ (Kennicutt 1998).  The
1.4-GHz radio emission was unresolved by the Very Large Array (VLA) in
its A configuration (Biggs \& Ivison 2006). A relatively
compact UV morphology is observed in the {\em Hubble Space Telescope}
imaging, with R$_{1/2}$=0.25\arcsec\ (Swinbank et al.\ 2006).

\smallskip
{\bf \rgb:}
In C04, the optical (rest-frame UV) spectrum was presented,
showing Ly$\alpha$ in emission but all other detectable lines in
absorption, and was classified as a pure starburst.  The near-IR
spectrum of \rgb\ from Keck/NIRSPEC covering the H$\alpha$/[N\,{\sc
ii}] region (Swinbank et al.\ 2004), again finds a large
H$\alpha$/[N\,{\sc ii}] ratio suggestive of a low metallicity,
[Fe/H]~$\sim$~$-$0.9.  The line width of the H$\alpha$ emission is
$450\pm 220$\,km\,s$^{-1}$ and the integrated line flux of
$2.3\pm1.2\times10^{-19}$\,W\,m$^{-2}$ then suggests an SFR uncorrected
for dust extinction of $110\pm 40$\,M$_\odot$\,yr$^{-1}$.  The radio
emission was unresolved by the VLA.  Only ground-based imaging exists
for this SFRG, showing a faint, unresolved source in 0.8$''$ seeing.

\smallskip
{\bf \rgc:}
An 
X-ray-detection and obscured AGN classification due to its X-ray
luminosity by Alexander et al.\ (2005) contrasts a PAH-dominated
mid-IR spectrum obtained by Pope et al.\ (2008), showing no obvious AGN
component.  No rest-frame UV spectrum for \rgc\ has been published as
there are no detected features. The redshift is based entirely on the
near-IR spectrum where \rgc\ was detected with Keck/NIRSPEC (Swinbank
et al.\ 2004), once again finding a large H$\alpha$/[N\,{\sc ii}]
ratio suggestive of an [Fe/H]~$\sim$~$-$0.8.  The narrow line width of
the H$\alpha$ emission ($112\pm 45$\,km\,s$^{-1}$) and the integrated
line flux of $0.4\pm0.3\times10^{-19}$\,W\,m$^{-2}$ suggests an SFR
uncorrected for dust extinction of $16\pm 9$\,M$_\odot$\,yr$^{-1}$.
Radio imaging of this SFRG with MERLIN (0.3\arcsec\ synthesized beam;
Casey et al.\ 2008b) reveals a double source structure with a
$\sim$1\arcsec\ elongated feature and a relatively compact
R$_{1/2}$=0.4\arcsec\ component.

\section{Results}

The velocity-integrated line fluxes or limits for all three SFRGs are listed in Table~1.
For \rga, inspection of the data cube shows a significant 4.9$\sigma$ detection of CO(3--2) line
emission at the phase center, integrated over the velocity channels at
$\sim -1000$\,km\,s$^{-1}$ with a velocity width of 400\,km\,s$^{-1}$
{\sc fwhm}.  Fitting a Gaussian profile to the CO spectrum, we derive
a best-fit redshift for the CO(3--2) emission of $z=2.1859\pm0.0002$, and 
estimate the CO flux is by summing the channels from $-2\sigma$ to $+2\sigma$
of Gaussian fit to the line.
We note that no significant continuum
emission is detected from the line-free region ($\sim$650\,MHz of
bandwidth) down to a 1$\sigma$ sensitivity of 0.07\,mJy\,beam$^{-1}$,
consistent with the submm limit, assuming a dust spectral index,
$\nu^{+3.5}$, for a modified blackbody with emissivity, $\beta=+1.5$.
For \rgb\ no significant emission is observed at the phase center,
although the limit on the CO gas mass is still of great interest relative to
the SMGs.
For \rgc\ we tentatively detected (3.2$\sigma$) a positive signal
integrated over the full band from $-600$ to $+600$\,km\,s$^{-1}$,
centered on the H$\alpha$ determined redshift of $z=1.996$.  A precise
CO redshift cannot be determined for \rgc\ as the line shape cannot be
determined in the low S/N spectrum.
To assess the possibility that we have simply detected continuum in this source, we analyze the radio spectral index, measured to be steep from the 8.4\,GHz/1.4\,GHz flux ratio (Muxlow et al.\ 2005),
S$_\nu \propto \nu^{-0.69}$, and the synchrotron contribution at $\sim$3\,mm is
negligible.  This does not however preclude a
contribution from an AGN component with the opposite spectral slope
emerging at higher frequencies. It is clear that the full radio spectrum
is needed to explore this issue, as well as the possibility of an obscured AGN.

In \rga, the CO-inferred redshift is close to the redshift measured
from various interstellar absorption lines in the Keck/LRIS UV
spectrum, but is blue shifted by 1100\,km\,s$^{-1}$ from the H$\alpha$
line detected by Swinbank et al.\ (2006).  Re-analysis of the near-IR
spectrum does not significantly change the result as the sky line
calibrations appear to be as presented in Swinbank et al.\ (2006).  If
the detected line were dominated by [N\,{\sc ii}] with
H$\alpha$/[N\,{\sc ii}]$<1$ -- and we stress that there is no evidence
in the spectrum of this -- then the implied velocity offset would be
$\sim$800\,km\,s$^{-1}$, somewhat closer to the average CO velocity
and consistent with the higher velocity peak in the detected CO
profile.  We cannot attribute the CO emission to an offset companion
(as was the case for SMG\,J09431 in Tacconi et al.\ 2006) as the CO
centroid is exactly at the near-IR and radio position to within the
1\arcsec\ centroiding uncertainty ($\sim$beam size $\times$
(S/N)$^{-1}$).  The H$\alpha$-inferred redshift has not always been
representative of the CO redshift in SMG surveys (the average
CO--H$\alpha$ offset is 150\,km\,s$^{-1}$), presumably because the
luminous core is so deeply dust enshrouded that wind outflows or
satellite H\,{\sc ii} regions are more strongly detected in H$\alpha$.
We therefore put forward the hypothesis that we are detecting a highly
dust-obscured gas-rich galaxy in CO(3--2), either a companion seen in
projection or else one not well sampled by the H$\alpha$ observations.
The rest-frame {\sc fwhm} of the CO line is $376\pm40$\,km\,s$^{-1}$,
close to that found in H$\alpha$ by Swinbank et al.\ (2006), but likely
a coincidence given that the CO and H$\alpha$ redshifts are discrepant.  

The 350$\mu$m SHARC-2 imaging of \rga\ shows a tentative continuum detection,
$S_{\rm 350\mu
m}=2.4\pm6.5$\,mJy at the radio position, however given the telescope pointing errors and
the low signal-to-noise (S/N) of any expected emission, we search a
region comparable to the beam size (9\arcsec). A 2$\sigma$ peak ($S_{\rm 350\mu
m}=13.7\pm6.9$~mJy) lies 7\arcsec\ from the radio position at 16h 36m
54.6s, +41$^{\circ}$ 04$'$ 28$''$ J2000.  Within this area there are
only $\sim$3 SHARC-2 beams and the chance of a spurious 2$\sigma$ peak
is only $\sim$6\%. There is a high likelihood ($\sim$90\%) that this
peak is related to RG\,J163655 and this flux range is completely
consistent with our SED fit to the radio photometry for \rga\ (Fig.~2).

We calculate from the integrated CO(3-2) intensity (Jy\,km\,s$^{-1}$) 
the line luminosities and estimate the total cold gas masses (H$_2$ + He) (listed in Table~1).
We adopt $L'_{CO(3-2)}=3.25\times10^7 S_{CO(3-2)} \nu_{obs}^{-2} (1+z)^{-3} D_L^2$, 
consistent with Solomon \& Vanden Bout (2005).

We assume both a line luminosity ratio of
$r_{32} = \frac{L'_{\rm CO(3-2)}}{L'_{\rm CO(1-0)}} = 1$ (i.e.\ a
constant brightness temperature) and a CO-to-H$_2$ conversion factor
of $\alpha = 0.8$\,M$_\odot$\,K\,km\,s$^{-1}$\,pc$^2$.  These values
are appropriate for local galaxy populations exhibiting similar levels
of star-formation activity to our SFRGs (e.g., local ULIRGs -- Solomon
et al.\ 1997), and this choice also facilitates comparison with SMGs modeled with the same values (Greve et al.\ 2005). We discuss later how the affect of
adopting typical Milky Way values $\alpha_{CO}$ and r$_{32}$.  


All three SFRGs show a
peaked SED in the mid-IR (Fig.~2) suggesting that this spectral region
is dominated by stars rather than AGN.  These properties are used to
derive the rest-frame $K$-band ($\sim2.2\mu$m) flux, and convert to a
stellar mass, in a matter similar to Borys et al.\ (2005), adopting
their L$_K$/M$_\odot = 3.2$ characteristic of a burst with an age of
$\sim250$\,Myr.  We interpolated between IRAC bands to estimate $S_{\rm
2.2 \mu m}$ (Table~3). 

\subsection{Derived Properties}
 
We then proceed to estimate various derived properties (listed in Table~3).
Starting with the gas surface density, for \rgc, we assume the CO emission traces the same large, extended morphology ($>1$\arcsec\ FWHM diameter) traced by resolved
MERLIN radio imaging (see Casey et al.\ 2008b for details), suggesting a low gas surface density. 
For \rga\ and \rgb, neither
the CO emission nor the VLA radio emission is resolved.
Without further information beyond the optical imaging described
previously, we assume the gas in these two SFRGs is distributed in a
disk with a similar radius to SMGs (e.g., Tacconi et al.\ 2008) of R$_{1/2}$=1.7\,kpc (0.25\arcsec), resulting in higher inferred gas surface densities
The CO luminosities for the three SFRGs are plotted as a function of FIR
luminosity and compared to other high-redshift galaxies detected in CO
in Fig.~3.

A dynamical mass for the well-detected \rga\ can be estimated by
analyzing the CO line profile.
We base our analysis on a single Gaussian fit to the line.  CO
emission is comparatively immune to the effects of obscuration and
outflows and therefore provides a unbiased measurement of dynamics
within the CO-emitting region.
The line width of the CO emission ($410\pm40$\,km\,s$^{-1}$ implies a
dynamical mass of ($8.4\pm 2.1) \times 10^{10} \csc^2 i$\,M$_\odot$,
assuming the gas lies in a disk with inclination $i$ and a radius of
0.25\arcsec\ (1.7\,kpc). Based on this, we calculate a gas to
dynamical mass fraction of $f = \frac{\mathrm{M_{\rm
gas}}}{\mathrm{M_{\rm dyn}}}$ $\sim 0.10\sin^2 i$.  We note that the
mean angle of randomly oriented disks with respect to the sky plane in
three dimensions is $i$ = 30$^{\circ}$ (Carilli \& Wang 2006),
resulting in an average inclination correction of $\csc^2 i = 4$.

\section{Analysis}
\subsection{Star-Formation Rate and Efficiency}

The radio luminosity of the SFRGs forms our baseline estimate for the
far-IR luminosity and SFRs (listed in Table~3), since we have only upper limits at 450 and
850\,$\mu$m.  
We caution however that our flux-limited
radio selection biases our sample to find objects of the same radio
luminosity as SMGs, regardless of the origin of the radio power. There
are clear examples within the wider SFRG sample where AGN dominate the
radio power despite an apparent starburst spectrum in the rest-frame
UV (e.g., Casey et al.\ 2008a).
The average $<{\rm SFR}_{\rm radio}>$ = 970\,M$_\odot$\,yr$^{-1}$, assuming the
radio/FIR relation $q= \log({\rm FIR}/3.75 \times 10^{12} {\rm Hz})/
S_{\rm 1.4GHz})$ (Helou et al.\ 1985), with $q=2.34$ (Yun et al.\
2001), a correction factor of 2.3 to total IR luminosity (TIR)
appropriate for hotter dust SEDs (Dale \& Helou 2002), and the
conversion from Kennicutt (1998)
$${\rm SFR} (M_\odot\,{\rm yr}^{-1}) = 1.8\times10^{-10} L_{\rm 8-1000
\mu m} (L_\odot).$$
The SFRs from their dust-corrected rest-frame 1500\AA\
continuum flux are factors $\sim50\times$ less (as described in C04),
and the UV emission is clearly not probing the true luminosities of
these systems (Table~3).
The H$\alpha$ emission line suggests SFRs (Table~3) ten times less than the radio
($<{\rm SFR}_{\rm H\alpha}>=92$), although with the average extinction
factor of $A_V \sim 2.9\pm0.5$ proposed for SMGs in Takata et al.\
(2006), this becomes  
$<{\rm SFR}_{\rm H\alpha, corr}>$ = 1300\,M$_\odot$\,yr$^{-1}$. 
This is consistent, on average, with the radio-inferred SFRs, although the
individual radio/H$\alpha_{\rm corr}$ ratios 
show very poor correspondence.  We have, of course, argued in the case of
\rga\ that the CO(3--2) and the H$\alpha$ line emission may be coming
from distinct regions, so these arguments do not obviously apply in
every case, and the average correction factor applied to the L$_{\rm
H\alpha}$ may not be appropriate either individually or for the population.
The 24-$\mu$m fluxes (Table~2) would represent strong supporting evidence for
large SFRs. We calculate SFR$_{24\mu m}$  as in Pope et al.\ (2006) for consistency with SMGs
(although strong 24-$\mu$m luminosities could also reveal a
dominant hot AGN dust torus). In a large sample of SFRGs, the 24-$\mu$m luminosity distribution
is indistinguishable from that of SMGs (Casey et al., 2008b). 
Two of our present three SFRGs have 24-$\mu$m observations, and only one is detected, 
the limit in the second case not being particularly constraining relative to the radio.
The SFR$_{\rm 24 \mu m}$ (Table~3) could be consistent with the SFR$_{\rm radio}$ given uncertainties in calibrating the SFR indicators.

For a reference point -- which can be scaled through by uncertainties
in the SFR -- we adopt the SFR$_{\rm radio}$ and calculate
surface densities $\Sigma_{\rm SFR}$, and star-formation efficiencies
(SFE = L$_{\rm FIR}/$M$_\mathrm{H_{2}}$) for the SFRGs, using assumed sizes described previously, and listed in Table~3. 
We find large SFEs ($4\times$ larger than the SMGs on average), although the
observational constraints on the SFRs for the SFRGs are consistent with having been over-estimated  by a factor $\sim2-3\times$ which would bring them into reasonable agreement
with the envelope of star-formation efficiencies for SMGs, LBGs and
local ULIRGs.

We can also roughly estimate a lower limit to the gas to dust mass ratio where CO is detected (Table~3).
We estimate similar dust mass limits for the SFRGs using
$$M_{\rm dust}[M_\odot] = \frac{1}{1+z}\frac{S_{obs}D_L^2}{\kappa_d B(\nu_0, T_d)}$$ 
assuming
$\kappa_{\nu} \propto \nu^\beta$, $\beta=+1.5$, and $B_\nu(T_{\rm d})
\sim \nu^+2$ in the M$_\mathrm{dust} < 1.6\times10^{8}$\,M$_\odot$ from the 2$\sigma$ limit on the
$850\,\mathrm{\mu m}$ flux of $\sim2.2$\,mJy, assuming a dust mass
absorption coefficient of $\kappa_{850_\mathrm{\mu
m}}=0.15\,\mathrm{m}^{2}\,\mathrm{kg}^{-1}$.  
We find ratios of $50$ and $100$, with at least a factor of $\sim6$
uncertainty accounting for our uncertainty of the dust temperature
($\Delta\,T_\mathrm{d} \simeq\: \pm\:{5}$\,K), dust emissivity
coefficient ($\Delta\,\beta \simeq\pm\:{0.5}$), and mass absorption
coefficient (about a factor of $\sim3$; e.g.~\citealt{Seaquist04}).

Assuming the molecular gas reservoirs we detect are fueling the star
formation within these galaxies, then there is enough gas to sustain
the current star formation for $\tau_{\rm depletion}\sim
M(\mathrm{H}_{2})/\mathrm{SFR}$ ranging from less than $9$ to 13\,Myr.
Since we have also estimated stellar masses, we can compare the gas
depletion time with the time to form the current stellar mass of the
system.  At the current SFRs,
$\tau_{\rm formation}\sim$M$_{\rm stars}/\mathrm{SFR}$ ranges from 80--240\,Myr,
which are comparable to the assumed ages of the stellar populations.

\begin{figure}
\centering
\includegraphics[width=8cm]{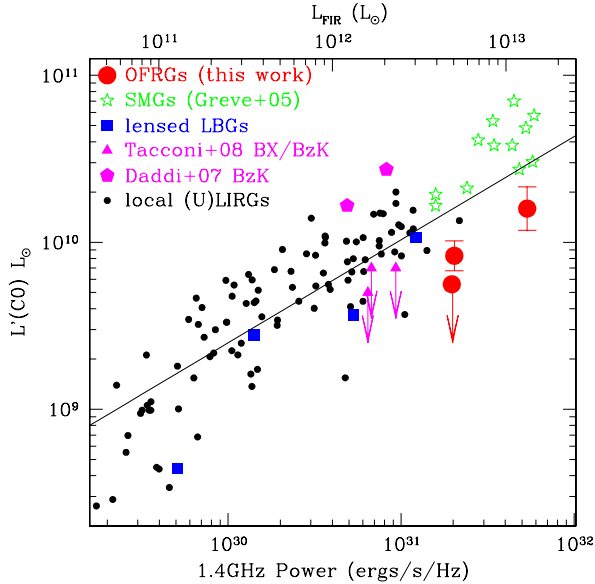}
\caption{A comparison of the CO and radio luminosities for the three
SFRGs described here, the SMGs from Greve et al.\ (2005) and Tacconi et
al.\ (2006), the lensed LBGs detected in CO (Baker et al.\ 2004,
Coppin et al.\ 2007, Kneib et al.\ 2005), the luminous $z\sim2$ BX/BzK
galaxies undetected in CO (Tacconi et al.\ 2008) and the two
$z\sim1.4$ BzK galaxies detected in CO by Daddi et al.\ (2008). 
The solid line is the best-fitting relation with a form of log L$'_{\rm
CO}=\alpha\,\mathrm{log}$\,L$_\mathrm{FIR}+\beta$ to the local LIRGs
and ULIRGs and the high-redshift SMGs from \citet{Greve05} (however it is no longer quite the best
fit when radio luminosity is considered consistently across the populations).  With gas
conversions fixed, the SFRGs appear to have lower CO gas masses than
SMGs, although if the radio-inferred SFRs are overestimated then the
SFRGs could still lie on the plotted gas/SFR relation.}
\label{fig:XCO}
\end{figure}

\begin{table*}
\begin{center}
\caption{Properties of CO observed SFRGs}
\label{tableSat}
\begin{tabular}{llcccccccccc}
\hline
{RA} & {Dec} & z(CO) & z(UV) & z(H$\alpha$) & L$_{\rm FIR, radio}$ &T$_{\rm d}$& S$_{CO}$ & L'$_{CO}$ & $FWHM_{CO}$  & M$_{\rm dyn}$ & M$_{\rm gas}$ \cr
{} & &  &  && 10$^{12}$ L$_\odot$ & {K} & ${\rm Jy\ km\, s^{-1} }$ & $10^{10} {\rm K\ km\,s^{-1}\ pc^{2}}$ & ${\rm km\,s^{-1}}$  & {10$^{10}$ M$_{\odot}$} & 10$^{9}$ M$_{\odot}$\cr
\hline
 16 36 55.04 & 41 04 32.0 & 2.189 & 2.186 & 2.192 &7.1$\pm$0.9 &$>46$ & $0.34\pm0.07$ &1.04$\pm$0.21 & 410$\pm$40 &8.4 $\sin^2 i$ & 8.3$\pm$1.7 \cr
 13 12 36.01 &  42 40 44.1 & {...} & 2.240 & 2.243 & 6.7$\pm$0.6 & $>48$ & $<0.27$\,$^a$ & $<$0.71 & {...} & {...}  & $<$5.6\cr
 12 37 11.34 &  62 32 31.0 & $\sim$1.996 & {...} & 1.996 & 16.7$\pm$0.7 & $>45$ & $0.70\pm
0.22$ & 1.93$\pm$0.60 & {...} & {...} & 15.4$\pm$4.8\cr
\hline
\end{tabular}
\end{center}
a) For \rgb, we set a limit for a 500 km\,s$^{-1}$ {\sc fwhm} line
centered at the H$\alpha$ redshift.\\
\end{table*}

\subsection{Comparison to other populations}

Comparison of the SFRGs to the SMGs is of primary importance, since a
major goal of the observations is to understand the degree to which
SFRGs should be treated on a similar footing to SMGs in models and
evolutionary calculations.  Taking the average CO line luminosity and
gas mass we find intrinsic line luminosities a factor of $\sim$4 lower
than the median for SMGs (cf.\
$<\!L'_\mathrm{CO}\!>=(3.8\,\pm{2.3})\times10^{10}$\ L$_\odot$ and
$<\!M_\mathrm{gas}\!>=(3.0\,\pm{1.6})\times10^{10}$\ M$_\odot$;
\citealt{Greve05}).  The CO line width of \rga\ is also much lower
than the median of SMGs ($<{\rm FWHM}>=780$\,km\,s$^{-1}$,
\citealt{Greve05}). 
Given the CO-inferred gas masses are somewhat low compared to SMGs we find, not
surprisingly, that the gas depletion timescales are short compared to
SMGs (which have $\tau_{\rm depletion}\sim 40$~Myr, Greve et al.\ 2005), though the
SFRGs are still within a physically plausible range.
We only have a useful constraint on the size of
the emitting region for \rgc\ to compare with higher resolution images
of SMGs, although in general the SFRGs exhibit similar radio sizes and
morphologies to SMGs (compare Chapman et al.\ 2004b and Biggs \&
Ivison 2008 to Casey et al.\ 2008b).  The typical gas to dynamical
mass fraction in SMGs is estimated to be $\sim0.3$ assuming a merger
model \citep{Greve05}, while they have SFEs of $L_{\rm
FIR}/M_\mathrm{H_2}\sim 450\pm 170$\,L$_\odot$\,M$_\odot^{-1}$
\citep{Greve05}, gas-to-dust mass ratios of $\sim200$ (with a factor
of a few uncertainty in the dust mass alone) and gas surface densities
of $\sigma_{\rm gas}\sim3000$\,M$_{\odot}$\,yr$^{-1}$\,pc$^{2}$
\citep{Tacconi06}.  Borys et al.\ (2005) estimate the average stellar
mass for SMGs at $z>1.5$ to be
$3.2^{+3.4}_{-1.6}\times10^{11}$~M$_\odot$, slightly larger than the
average $1.8\times10^{11}$~M$_\odot$ for our three SFRGs.

Overall, this comparison suggests that SFRGs may be somewhat
smaller mass objects (lower stellar mass, lower CO mass and lower
dynamical mass) than SMGs, but share with them a large radio luminosity. 
The SFRs of both SMGs and SFRGs are
subject to sizable uncertainties, not least of which is the initial
mass function (e.g., Baugh et al.\ 2005).  Pope et al.\ (2006) have
pointed out that SMG SFRs estimated from the 24\,$\mu$m {\it
Spitzer}-MIPS observations are lower than those estimated from the
850\,$\mu$m or radio wavebands, although this could represent an issue
of relative calibrations of these indicators in this luminosity regime rather than intrinsic properties. 

We can also compare the SFRGs to local populations and less luminous
star-forming galaxies at $z\sim2$.  Locally, L$'_{\rm CO}$ increases
with L$_{\rm FIR}$ for (U)LIRGs, with the \citet{Greve05} sample of
SMGs extending this trend out to the highest far-IR luminosities
($\gtrsim10^{13}$L$_{\odot}$).  For comparison, in Fig.~3 we have
plotted the SFRGs on the L$'_{\rm CO}$--L$_{\rm FIR}$ diagram along
with SMGs lying on the local relation, as well as three LBGs from the
literature within the considerable uncertainties in their far-IR
luminosities, three undetected BX/BzK galaxies (Tacconi et al.\ 2008),
and two CO-detected BzK galaxies (Daddi et al.\ 2008 -- these sources
lie above the relation).  The SFRGs lie somewhat below this relation.

\begin{table*}
\begin{center}
\caption{Photometric Properties of  SFRGs}
\label{tableSat}
\begin{tabular}{lcccccccccc}
\hline
{source} & 20cm & 850$\mu$m&350$\mu$m &24$\mu$m &8.0$\mu$m & 5.8$\mu$m & 4.5$\mu$m & 3.6$\mu$m& $I$ & $R$\\ 
{} & $\mu$Jy & mJy & mJy &$\mu$Jy &$\mu$Jy  & $\mu$Jy & $\mu$Jy  & $\mu$Jy & ${AB}$\,mag &${AB}$\,mag \\
\hline
RG\,J163655 & 43.9$\pm$7.1 & -1.5$\pm$1.1 & 13.7$\pm$6.9 & $<150$ & 14.4$\pm$2.7 & 16.8$\pm$3.5 & 11.9$\pm$1.1 & 9.9$\pm$0.7 &23.5 & 23.8 \cr
RG\,J131236 & 48.7$\pm$4.1 & 0.4$\pm$1.1 & {...} & {...} & 8.4$\pm$2.7 & $<11$ & 7.0$\pm$1.1 & 5.4$\pm$0.7   & 24.2 &25.0 \cr
RG\,J123711 & 133.1$\pm$5.1 & 2.2$\pm$1.2 & {...}  & 531$\pm8$ & 37.5$\pm$0.8 & 42.1$\pm$1.2 & 35.8$\pm$0.3& 32.3$\pm$0.2 & 24.9 & 25.8 \cr
\hline
\end{tabular}
\end{center}
\end{table*}

\begin{table*}
\begin{center}
\caption{Derived properties of the SFRGs}
\label{tableSat}
\begin{tabular}{lcccccccccc}
\hline
{source}  & SFR$_{radio}$ & SFR$_{UV}$ & SFR$_{H\alpha}$ &SFR$_{24\mu m}$\,$^a$ & SFE\,$^b$ & $\Sigma_{gas}$ & M$^*$\,$^c$ & gas/dust\,$^d$  & $\tau_{depletion}$ & $\tau_{gas\, to\, stars}$ \cr
{}  & M$_\odot$\,yr$^{-1}$ & M$_\odot$\,yr$^{-1}$  & M$_\odot$\,yr$^{-1}$ & M$_\odot$\,yr$^{-1}$  &L$_\odot$/M$_\odot$ & {M$_\odot$\,yr$^{-1}$\,kpc$^{-2}$} & M$_\odot$  & {}  & (Myr) &  (Myr) \cr
\hline
 RG\,J163655 & $630\pm120$ & $46\pm11$ & $150\pm50$ & $<255$  & $600$ & $70$ &  $1\times10^{11}$ & 52 & 13.2 &160 \cr
 RG\,J131236 &  $610\pm91$ & $33\pm16$ & $110\pm40$ & {--} & $<1450$ & $<65$ &$5\times10^{10}$ & --  & $<9.2$  & 80 \cr
 RG\,J123711 & $1670\pm113$ &$11\pm10$& $16\pm9$ & $880\pm54$  & $1330$ & $35$ & $4\times10^{11}$ & 100 & 9.2 & 240 \cr
\hline
\end{tabular}
\end{center}
$a)$ SFR from 24$\mu$m luminosity, 
assuming the average SED in Dale \& Helou (2002).\\
$b)$ Star formation efficiency (SFE) is the SFR divided by the molecular gas mass.\\
$c)$ Stellar mass calculated as Borys et al.\ (2005), adopting
their L$_K$/M$_\odot = 3.2$.\\
$d)$ No limit is possible for \rgb\ since both gas and dust are limits.\\
\end{table*}

\section{Discussion}

As our calculations above have shown, it is very difficult to estimate
the precise gas masses for SFRGs due to various uncertainties.
However, a potentially important result emerges from these CO observations: compared
to SMGs, SFRGs appear to be significantly more efficient at producing
stars from a given molecular gas mass.  If this is strictly true, then
SFRGs cannot be interpreted as {\em scaled up} versions of local ULIRGs
as Tacconi et al.\ (2006, 2008) have argued is the case for SMGs,
since their gas masses appear to be lower than expected for their
radio luminosities.
There are two considerations to be taken into account here. Firstly,
the far-IR luminosities and thus SFRs may be over-estimated from the
radio, for instance if buried AGN were present.  While none of these
sources have AGN signatures in their UV or optical spectra, a deeply
obscured AGN could still be driving a significant portion of the radio
luminosity (e.g., Daddi et al.\ 2007b; Casey et al.\ 2008a).  
A further possible complication comes again from the radio selection of these objects.
There is a $\sim$0.25~dex scatter in the radio-FIR relation (Yun et al.\ 2001), and while locally there is no apparent correlation between SED shape and radio-FIR scaling, it is possible in our SFRGs that we are selecting galaxies which are amongst the lower 0.25~dex (weaker FIR per unit radio).
If the SFRs in our galaxies were several times lower then the efficiencies would be similar
to the average SMG (Fig.~3).
Secondly, the conversion from CO(3--2) to molecular gas mass may not
be the same as for SMGs. If $\alpha_{\rm CO}$ were greater than the $\sim$1
estimated for local ULIRGs (and as inferred to be correct for SMGs),
these SFRGs could make up the shortfall in molecular gas mass from the
average SMG ($\sim4\times$), although the conversion would have to
approach that typically adopted for the Milky Way $\alpha_{\rm CO}=4.6$
(Solomon \& van den Bout 2005). This is unlikely given that the SFRGs
often show clear evidence in high-resolution radio observations for
merger-driven starbursts (Casey et al.\ 2008b).

It is noteworthy that while our observations have highlighted an
ultraluminous $z\sim2$ population which may build stars in an
extremely efficient mode (or at least as efficient as SMGs if their
SFRs are overestimated by factors of several), recent observations
(Daddi et al.\ 2008) have identified a population of $z\sim1.5$
galaxies which they detect in CO(2--1) exhibiting the opposite
property: low-efficiency star formation. These Daddi et al.\ galaxies
are selected as large massive disk-like galaxies, and it is perhaps
not surprising that they form stars in an apparently quiescent
``spiral-galaxy'' mode.  Nonetheless, it is intriguing that galaxies
in the high-redshift Universe have been discovered with such a wide
range of star-forming efficiencies (from poor to extreme) all lying in
the 10$^{12-13}$-L$_\odot$ regime.  Massive galaxies are being built
in a variety of modes in the $z=1-3$ peak star-formation period.

The short depletion times compared to the long times to form the stellar masses  
suggests we may be seeing
SFRGs in the last phase of their current star-formation episodes.
However, one would expect on average to find SFRGs as a population
half-way through their gas consumption lifetimes. In this context, the
small ages imply either a very high duty cycle or that the SFRs are
over-estimated from the radio luminosity.
We reiterate that the large FIR luminosity estimates for our SFRGs are
based mainly on the radio/FIR relation which is apparently applies
for SMGs (Kovacs et al.\ 2006), but is only marginally supported
for SFRGs through the 24-$\mu$m luminosity and dust-corrected H$\alpha$
measurements.  It is possible that we are over-estimating their
star-formation activity, which would lead us to under-estimate the
duration timescales above.

We also note that the observed CO may be closely
associated with the star formation, and thus is probably warm and highly visible
(i.e. with low $\alpha_{CO}$). It does not rule out the existence of a
cooler less visible component (with high $\alpha_{CO}$), not intimately
associated with the current zone of star formation (e.g.\ in the inner disk), which
may still become available within its own dynamical timescale to fuel star formation.
The result above may therefore be affected by the selection for
the highly visible component of the molecular gas.
However, the cold gas would need to be fairly widely distributed (to ensure that it  
doesn't violate the dynamical limits on the total gas+stellar mass in  
the central regions), yet it must also be able to flow into the  
central regions on a timescale of $<$100~Myrs (near the limit of the gas sound speed $\sim100$~km/s).

In order to evaluate the implications of these results for massive
galaxy formation, we recall that these three galaxies have typical
characteristics for the SFRG population at all wavelengths measured.
Our results underline the major role of gas consumption over short
timescales and with high efficiencies, characterizing rapid and strong
merger-driven bursts as a major growth mode for both stellar mass and
black holes in the distant Universe.
Even if the SFRs in these SFRGs were over-estimated by a factor of a
few, they would remain ULIRG-class galaxies. If we assume that
$\sim$50\% of the submm-faint SFRGs at $z\sim2$ are dominated by star
formation at levels comparable to SMGs, we arrive at a density of
$\sim5\times10^{-6}$\,Mpc$^{-3}$, similar to that observed for SMGs
(Chapman et al.\ 2003).  Together, the SMGs and SFRGs represent a
volume density $10\times$ smaller than measured for galaxies inferred
to be forming stars at low efficiencies by Daddi et al.\ (2007b) which
have space densities of order of $10^{-4}$\,Mpc$^{-3}$.  With SFRs a
few to 10$\times$ larger in the SMGs and SFRGs, the net effect is
roughly equal numbers of stars being formed in both high- and
low-efficiency modes at $z\sim2$.

Further study should ascertain whether other SFRGs follow a similar
pattern to the galaxies studied in this paper. A substantial sample
will allow the properties of the gas to dynamical mass ratio to be
determined accurately for the population, since we only
currently have one well-detected line profile, and the
dynamical mass determination is limited by the unknown inclination.

\section{Conclusions}

%
%

$\bullet$ We conclude that the radio luminosities of these SFRGs are higher for their overall mass (gas plus stellar) than for the SMGs (given that if the radio-SFRs are over-estimated for one class, they could well be for both). 

$\bullet$ We note that SMGs in general, and also these SFRGs, are
outliers of the stellar mass-SFR correlation (Daddi et al.\ 2007a),
probably due to the higher efficiency in forming stars for a similar
stellar mass and CO luminosity. Lower gas masses in the SFRGs would
imply even higher SFEs than the SMGs.  By contrast, if the SFRs are
significantly over-estimated by the radio, or the CO-to-H$_2$
conversion were significantly different from SMGs, then SFRGs could
have similar efficiencies to typical ULIRGs.
Together with the apparent low-efficiency star-forming (U)LIRGs from
Daddi et al.\ (2008), the SMGs and SFRGs with SFRs several to
10$\times$ larger than the Daddi et al.\ galaxies suggest roughly
equal numbers of stars being formed in both high- and low-efficiency
modes at $z\sim2$.
Massive galaxies are being built
in an impressive variety of modes in the $z=1-3$ peak star-formation period.

$\bullet$ If the radio-inferred SFRs are correct, then these SFRGs are more efficient star formers than SMGs, and 
cannot obviously be interpreted as {\em scaled up} versions of local
ULIRGs as Tacconi et al.\ (2006, 2008) have argued is the case for
SMGs.  The SFRGs' radio luminosities are larger than would naturally
scale from local ULIRGs given the gas masses or gas fractions.
These observed gas masses and star-formation properties may be typical
of the SFRG population and further work is justified to explore this
population with improved statistics.

$\bullet$ Our results underscore the fact that ultraluminous galaxies
in the high-redshift Universe have been discovered with a wide
range of star-forming efficiencies, the SFRGs apparently being one extreme.
Massive galaxies are likely being built
in a variety of modes in the $z=1-3$ peak star-formation period.

\section*{acknowledgements}

We thank an anonymous referee for a very careful reading and helpful comments.
This work is based on observations carried out with the IRAM Plateau de Bure
Interferometer. IRAM is supported by INSU/CNRS (France), MPG (Germany)
and IGN (Spain). SCC acknowledges a fellowship from the Canadian Space
Agency and an NSERC discovery grant.  IRS acknowledges support from
the Royal Society.  AMS acknowledges support from STFC.  We
acknowledge the use of GILDAS software
(http://www.iram.fr/IRAMFR/GILDAS).


\begin{thebibliography}{}

\bibitem[Barger et al.(2000)]{Barger00}
Barger, A., Cowie, L., \& Richards, E., 2000, \aj, 119, 2092
\bibitem[Baker et al.(2004)]{Baker04}
Baker, A.J., Tacconi, L.J., Genzel, R., Lehnert, D., \& Lutz, D., 2004, \apj, 604, 125
\bibitem[Baugh et al.(2005)]{Baugh05}
Baugh, C.M., Lacey, C.G., Frenk, C.S., Granato, G.L., Silva, L., Bressan,
Benson, A.J., \& Cole, S., 2005, \mnras, 356, 1191
\bibitem[Biggs \& Ivison(2006)]{Biggs06}
Biggs, A.D., Ivison, R.J, 2006, MNRAS, 371, 963
\bibitem[Biggs \& Ivison(2008)]{Biggs08}
Biggs A.D.,  Ivison R.J., 2008, MNRAS, 385, 893 
\bibitem[Blain et al.(2002)]{Blain02}
Blain, A.W., Smail, I., Ivison, R.J., Kneib, J.-P., \& Frayer, D.T.,
2002, Phys.\ Rep., 369, 111
\bibitem[Blain(1999)]{Blain99}
Blain, A.W. 1999, MNRAS, 309, 955
\bibitem[Borys et al.(2003)]{Borys03}
Borys C., Chapman S.C., Halpern M., Scott D.,
2003, MNRAS, 344, 385
\bibitem[Borys et al.(2005)]{Borys05}
Borys, C., Smail, I., Chapman, S.C., Blain, A.W., Alexander, D.M., \& Ivison, R.J.,
2005, \apj, 635, 853
\bibitem[Calzetti et al.(2000)]{Calzetti00}
Calzetti, D., Armus, L., Bohlin, R.C., Kinney, A.L., Koornneef, J., \&
Storchi-Bergmann, T., 2000, \apj, 533, 682
\bibitem[Carilli \& Wang(2006)]{CarWangErr}
Carilli, C.L., \& Wang, R., 2006, \aj, 132, 2231
\bibitem[Casey et al.(2008b)]{casey08b}
Casey, C., et al.\  MNRAS, 2008, submitted
\bibitem[Chapman et al.(2000)]{Chapman00}
Chapman, S.C., et al.\ 2000, \mnras, 319, 318
\bibitem[Chapman et al.(2001)]{chapman01}
Chapman, S.C., Richards, E., Lewis, G., Wilson, G., \& Barger, A., 2001, ApJ, 548, L147 
\bibitem[Chapman et al.(2002)]{chapman02}
Chapman, S.C., Lewis, G., Scott, D., Borys, C., \& Richards, E., 2002, ApJ, 570, 557 
\bibitem[Chapman et al.(2003a)]{chapman03a}
Chapman S., Blain A., Ivison R., Smail I., 2003, Nat, 422, 695 
\bibitem[Chapman et al.(2003b)]{chapman03b}
Chapman, S.C., Barger, A., Cowie, L., Scott, D., Borys, C., Lewis, G.,
Richards, E, Steffan, A., \& Wilson, G., 2003, ApJ, 585, 57 
\bibitem[Chapman et al.(2004a)]{chapman04a}
Chapman S.C., Smail I., Blain A.,  Ivison R., 2004, ApJ, 614, 671 
\bibitem[Chapman et al.(2004b)]{chapman04b}
Chapman S.C., Smail, I., Windhorst, R., Muxlow, T., \& Ivison, R.J.,
2004, ApJ, 611, 732 
\bibitem[Chapman et al.(2005)]{chapman05}
Chapman S.C., Blain A., Smail I.,  Ivison R., 2005, ApJ, 622, 772 
\bibitem[Daddi et al.(2007a)]{dad07}
Daddi, E., et al., 2007a, ApJ, 670, 156 
\bibitem[Daddi et al.(2007b)]{dad07b}
Daddi, E., et al.\ 2007b, ApJ, 670, 173 
\bibitem[Daddi et al.(2008)]{daddi08}
Daddi, E., et al.\ 2008, \apj, 673, L21
\bibitem[Dale \& Helou(2002)]{DaleHelou02}
Dale, D.A., \& Helou, G., 2002, \apj, 576, 159
\bibitem[Dale et al.(2001)]{Dale01}
Dale, D.A., Helou, G., Contursi, A., Silbermann, N.A., \&
Kolhatkar, S., 2001, \apj, 549, 215
\bibitem[Desai et al.(2006)]{desai06} 
Desai, V., et al.\ 2006, ApJ, 641, 133
\bibitem[Greve et al.(2005)]{Greve05}
Greve, T.R., et al.\ 2005, \mnras, 359, 1165
\bibitem[Hainline et al.(2008)]{Hainline08}
Hainline, L.J.\ et al.\ 2008, ApJ, in press
\bibitem[Helou(1985)]{Helou85}
Helou G., Soifer B., Rowan-Robinson, M., 1985, \apj, 298, L7
\bibitem[[Hinshaw et al.(2008)]{Hinshaw08}
Hinshaw, G., et al., 2008, arXiv0803.0732
\bibitem[[Houck et al.(2005)]{Houck05}
Houck, J., et al.\ 2005, ApJ, 622, 105
\bibitem[Hughes et al.(1998)]{Hughes}
Hughes, D.\,H., et al.\ 1998, Nat, 394, 241
\bibitem[Ivison et al.(2002)]{i02}
Ivison, R.J., et al.\ 2002, MNRAS, 337, 1
\bibitem[Ivison et al.(2007a)]{i07a}
Ivison, R.J., et al.\ 2007a, ApJ, 660, L77
\bibitem[Ivison et al.(2007b)]{i07b}
Ivison, R.J., et al.\ 2007b, MNRAS, 380, 199
\bibitem[Kennicutt(1998)]{Kennicutt98}
Kennicutt, R.C., Jr., 1998, \apj, 498, 541
\bibitem[Kneib et al.(2005)]{Kneib05}
Kneib, J.-P., Neri, R., Smail, I., Blain, A., Sheth, K., van der Werf, P.,
\& Knudsen, K.K., 2005, A\&A, 434, 819
\bibitem[Kovacs et al.(2006)]{kovacs06}
Kovacs, A., Chapman, S.C., Dowell, C.D., Blain, A.W., Ivison, R.J.,
Smail, I., \& Phillips, T.G.,  2006, ApJ, 650, 592 
\bibitem[Leitherer et al.(1999)]{leith99}
Leitherer, C., et al.\ 1999, ApJS, 123, 3
\bibitem[Menendez-Delmestre et al.(2007)]{karin07}
Menendez-Delmestre, K., et al., 2007, ApJ, 655, 65
\bibitem[Pope et al.(2006)]{pope06}
Pope, A., et al.\ 2006, MNRAS, 370, 1185
\bibitem[Pope et al.(2008)]{pope08}
Pope, A., et al.\ 2008, in press
\bibitem[Sajina et al.(2007)]{sajina07}
Sajina, A. et al.\ 2007 (astro-ph/0705.3377)
\bibitem[Scott et al.(2002)]{scott02}
Scott, S.E., et al.\ 2002, MNRAS, 331, 817
\bibitem[Seaquist et al.(2004)]{Seaquist04}
Seaquist E., Yao L., Dunne L., et al.\ 
2004, \mnras, 349, 1428
\bibitem[Smail, Ivison \& Blain(1997)]{SIB}
Smail, I., Ivison, R.J., \& Blain, A.W., 1997, \apj, 490, L7
\bibitem[Smail et al.(2007)]{Smail06}
Smail, I. et al.\ 2007, \apj, 654, 33
\bibitem[Solomon \& Vanden Bout(2005)]{Solomon_rev}
Solomon, P.M., \& Vanden Bout, P.A., 2005, \araa, 43, 677
\bibitem[Solomon et al.(1997)]{S097}
Solomon, P.M., Downes, D., Radford, S.J.E., \& Barrett, J.W., 1997, \apj, 478, 144
\bibitem[Spergel et al.(2003)]{Spergel}
Spergel, D.N., et al.\ 2003, \apjs, 148, 175
\bibitem[{{Steidel} {et~al.}(2004){Steidel}, {Shapley}, {Pettini},
  {Adelberger}, {Erb}, {Reddy}, \& {Hunt}}]{steidel04}
{Steidel}, C.~C., {Shapley}, A.~E., {Pettini}, M., {Adelberger}, K.~L., {Erb},
  D.~K., {Reddy}, N.~A., \& {Hunt}, M.~P. 2004, \apj, 604, 534
\bibitem[Swinbank et al. (2004)]{Swinbank04}
Swinbank, A.M., Smail, I., Chapman, S.C., Blain, A.W., Ivison, R.J., \&
Keel, W.C., 2004 \apj, 617, 64
\bibitem[Swinbank et al. (2005)]{Swinbank05}
Swinbank, A.M., 
Smail I., Bower R., Borys C., Chapman S.,  et al., 
2005, MNRAS, 359, 401
\bibitem[Swinbank et al. (2006)]{Swinbank06}
Swinbank, A.M., Chapman, S.C., Smail, I., Lindner, C., Borys, C.,
Blain, A.W., Ivison R.J., \& Lewis, G.F., 2006, \mnras, 371, 465
\bibitem[Tacconi et al.(2006)]{Tacconi06}
Tacconi, L.J., et al.\ 2006, \apj, 640, 228
\bibitem[Tacconi et al.(2008)]{Tacconi08}
Tacconi, L.J., et al.\ 2008, ApJ, in press
\bibitem[Valiante(2007)]{valiante}
Valiante, E., et al., 2007, ApJ, 660, 1060
\bibitem[Weedman et al.(2006a)]{weed06a}
Weedman, D., et al.\ 2006a, ApJ, 638, 613 
\bibitem[Weedman et al.(2006b)]{weed06b}
Weedman, D., et al.\  2006b, ApJ, 651, 101 
\bibitem[Yan et al.(2005)]{yan05}
Yan, et al.\ 2005, ApJ, 628, 604 
\bibitem[Yan et al.(2007)]{yan07}
Yan, et al.\ 2007, ApJ, 658, 778 
\bibitem[Yao et al.(2003)]{Yao03}
Yao, L., Seaquist E.R., Kuno N., \& Dunne L., 2003, \apj, 588, 771
\bibitem[Yun et al.(2001)]{Yun01}
Yun, M., Reddy, N., \& Condon, J., 2001, \apj, 554, 803

\end{thebibliography}
\end{document}